\documentclass[aps,prl,superscriptaddress,twocolumn]{revtex4-1}
\usepackage{amssymb}
\usepackage{bm}
\usepackage{graphicx}
\usepackage{color}
\usepackage[T1]{fontenc}

\usepackage{lipsum}

\makeatletter
\def\blfootnote{\xdef\@thefnmark{}\@footnotetext}
\makeatother
\usepackage[bottom]{footmisc}

\begin{document}
\title{Rayleigh-Taylor turbulence with singular non-uniform initial conditions}
\author{L. Biferale}
\affiliation{Department of Physics and INFN, University of Rome ``Tor Vergata'', Via della Ricerca Scientifica 1, 00133 Rome, Italy}
\author{G. Boffetta}
\affiliation{Department of Physics and INFN, University of Torino,  via P. Giuria 1, 10125 Torino, Italy}
\author{A.A. Mailybaev}
\affiliation{Instituto Nacional de Matem\'atica Pura e Aplicada -- IMPA, 22460-320 Rio de Janeiro, Brazil}
\author{A. Scagliarini}
\affiliation{Istituto per le Applicazioni del Calcolo 'M. Picone' -- IAC-CNR, Via dei Taurini 19, 00185 Rome, Italy}
\begin{abstract}
We perform Direct Numerical Simulations of three dimensional Rayleigh-Taylor turbulence with a non-uniform singular initial temperature background. In such conditions, the mixing layer evolves under the driving of a varying {\it effective}  Atwood number; the long time growth is still self-similar, but not anymore proportional to $t^2$ and depends on the singularity exponent $c$ of the initial profile $\Delta T \propto z^c$. We show that universality is recovered when looking at the {\it  efficiency}, defined as the ratio of the variation rates of the kinetic energy over the heat flux. A closure model is proposed that is able to reproduce analytically the time evolution of the mean temperature profiles, in excellent agreement with the numerical results. Finally, we reinterpret our findings on the light of {\it spontaneous stochasticity}  where the growth of the mixing layer is mapped in to the propagation of a wave of turbulent fluctuations on a rough background. 
\end{abstract}
\maketitle
%
%
%
\noindent {\sc Introduction.}
Turbulent mixing\blfootnote{Postprint version of the manuscript {\it Phys. Rev. Fluids} {\bf 3}, 092601(R) (2018).} is a mechanism of utmost importance in many natural and industrial processes, often induced by the Rayleigh-Taylor (RT) instability which takes place when a fluid is accelerated against a less dense one \cite{sharp1984overview,kull1991theory,abarzhi2010review,boffetta2017incompressible,zhou2017rayleigh1}. RT turbulence occurs in disciplines as diverse as in astrophysics \cite{bell2004direct,zingale2005three,cabot2006reynolds},  atmospheric science \cite{KelleyEtAl}, confined nuclear fusion \cite{Lindl,Atzeni},plasma physics \cite{takabe}, 
laser-matter interactions \cite{sgattoni,shigemori}
(see \cite{boffetta2017incompressible,zhou2017rayleigh1,zhou2017rayleigh2} for recent reviews).
One important application of RT instability is the case of convective flow, in which density differences reflect temperature fluctuations of a single fluid and the acceleration is provided by gravity.

In the simplest configuration of Boussinesq approximation for an incompressible flow, RT turbulence considers a planar interface which separates a layer of cooler (heavier, of density $\rho_H$) fluid over a layer of hotter (lighter, of density $\rho_L$) fluid under a constant body force such as gravity. 
The driving force is  constant in time and proportional to $g \mathcal{A}$, where $g$ is the acceleration due to the body force and $\mathcal{A} = (\rho_H - \rho_L)/(\rho_H+\rho_L)=\beta \theta_0/2$ is the Atwood number, expressed in term of the thermal expansion coefficient $\beta$ and the temperature jump $\theta_0$ between the two layers.
However, in some relevant circumstances one has to cope with time varying acceleration (as in inertial confinement fusion or in pulsating stars  \cite{dimonte1996turbulent,dimonte2007rayleigh,LivescuEtAlJOP})
or with a  varying Atwood number, that emerges when the mixing proceeds over a non-uniform background as in  thermally stratified atmosphere \cite{MoninObukhov,KaderYaglom,biferale2011pre}. 

In this work we address a question with both fundamental and applied importance: what happens when the initial unstable profile is more general than the usual RT step-function, as in the case of non-differentiable power-law density profile.
As a result, the mixing layer will evolve in a non-homogeneous background. In particular we investigate analytically and by using direct numerical simulations in three dimensions the generic case when the initial unstable vertical temperature distribution is given by the power law: 
\begin{equation}
T_0(z)=-(\theta_0/2)\,\mathrm{sgn}(z) \left(\frac{|z|}{L}\right)^{c},
\label{eq2.3}
\end{equation}
where $L$ is a characteristic length scale 
and $-L \le z \le L$.
The exponent of the singularity belongs to the interval $-1 < c < 1$, where the upper limit corresponds to a smooth profile and 
the lower limit ensures that the potential energy density,  $-\beta g z T_0(z)$, does not diverge near the interface among the two miscible fluids at  $z = 0$. The value $c=0$ recovers the standard RT configuration.

We develop a closure model based on the Prandtl Mixing Length approach, which is able to reproduce with good accuracy the evolution of the mean temperature profile at all scales and for all values of the singularity exponent $c$. Beside the importance of testing the robustness with respect to the initial configuration, the above setup allows us to investigate the idea that the Mixing Layer (ML) growth can be mapped to a traveling wave in appropriate renormalized variables. This wave describes the self-similar evolution of the probability distribution function (PDF) of turbulent fluctuations  from small to large scales in a rough background given by the initial singular profile ~\cite{leith,eyink}. Such a description would then naturally explain the universality of the ML evolution and its spontaneously stochastic behavior in the inertial range~\cite{mailybaev2017}. We introduce a shell model for the RT evolution to illustrate and quantify the ML statistical properties. 


\noindent {\sc Results for Navier--Stokes Equations.}
We consider the Boussinesq approximation for an incompressible 
velocity field ${\bf u}({\bf r},t)$ coupled to the temperature field $T({\bf r},t)$ by a buoyancy term:
\begin{eqnarray}
\partial_t{\bm u}+ {\bm u} \cdot {\bm \nabla} {\bm u} 
& = & - {\bm \nabla} p + \nu \nabla^2 {\bm u} - \beta {\bm g} T,
\label{eq2.1}\\
\partial_t T
+ {\bm u} \cdot {\bm \nabla} T & = &
\kappa \nabla^2 T,
\label{eq2.2}
\end{eqnarray}
where ${\bm g}=(0,0,-g)$ is the gravity acceleration, 
$\nu$ and $\kappa$ are the kinematic viscosity and thermal diffusivity 
respectively.
The choice to rely on Navier-Stokes-Boussinesq equations for studies of RT at high Reynolds numbers is very widespread, and it is justified by 
the observation that the turbulent Mach number has an upper bound \cite{Mellado}, thus making RT turbulence an effectively incompressible (or low compressible)
phenomenon. On the other hand, it is known that, when detectable, compressibility effects amount mainly to break the up-down symmetry of the mixing layer growth, with 'spikes' (downward falling temperature fluctuations) being on average faster than 'bubbles' (upward rising); nevertheless, such asymmetry is limited to the prefactor, while the scaling in time of the full mixing layer width, which is our main interest here, is preserved \cite{ScagliariniPOF,LivescuRistoJFM}.

The initial condition for the velocity at position ${\bf r} = (x,y,z)$ is ${\bm u}({\bm r},0)=0$, 
while for the temperature field $T({\bm r},0) = T_0(z)$ we consider a generic power-law
distribution given by (\ref{eq2.3}). The only inviscid parameter that relates  spatial and temporal scales is $\xi = \beta g\theta_0/L^c$ which has physical dimensions of $\left[\textrm{length}^{1-c}/\textrm{time}^2\right]$. Thus, for 
a given length $L$, the corresponding integral temporal scale is given by $t_* = \xi^{-1/2}L^{(1-c)/2} = \sqrt{L/(\beta g\theta_0)}$.
The distribution (\ref{eq2.3}) is unstable and the dimensional argument provides the inviscid growth exponent $\lambda \simeq \xi^{1/2} k^{(1-c)/2}$ for the modes with wavenumber $k$, where the dimensionless proportionality coefficient can be determined by solving the linear stability problem~\cite{chandrasekhar}. 
This dispersion relation predicts that the instability is driven by the smallest scales for all $c < 1$.

\begin{figure}[t]
\includegraphics[width=0.25\columnwidth]{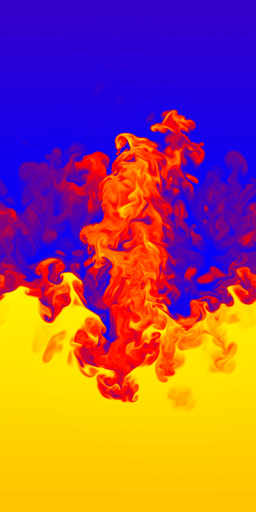}
\includegraphics[width=0.25\columnwidth]{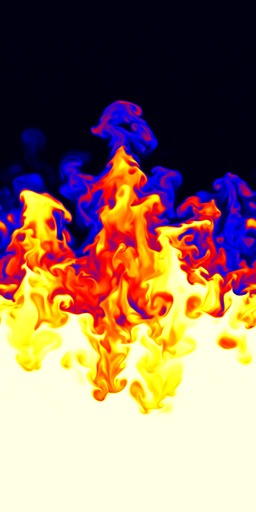}
\includegraphics[width=0.25\columnwidth]{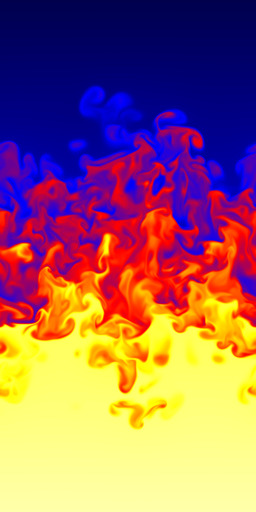}
\caption{(Color online). Snapshots of the vertical section of the temperature field $T$ for three simulations of RT turbulence with power law initial condition (\ref{eq2.3}) with $c=-0.25$ (left), $c=0$ (center) and $c=0.25$ (right) at three different times corresponding to the same mixing length
$h(t) \simeq 0.4 L_z$. High (low) temperature is represented by yellow (blue).}

\label{fig:snapshot}
\end{figure}
\begin{figure}[t]
\includegraphics[width=1\columnwidth]{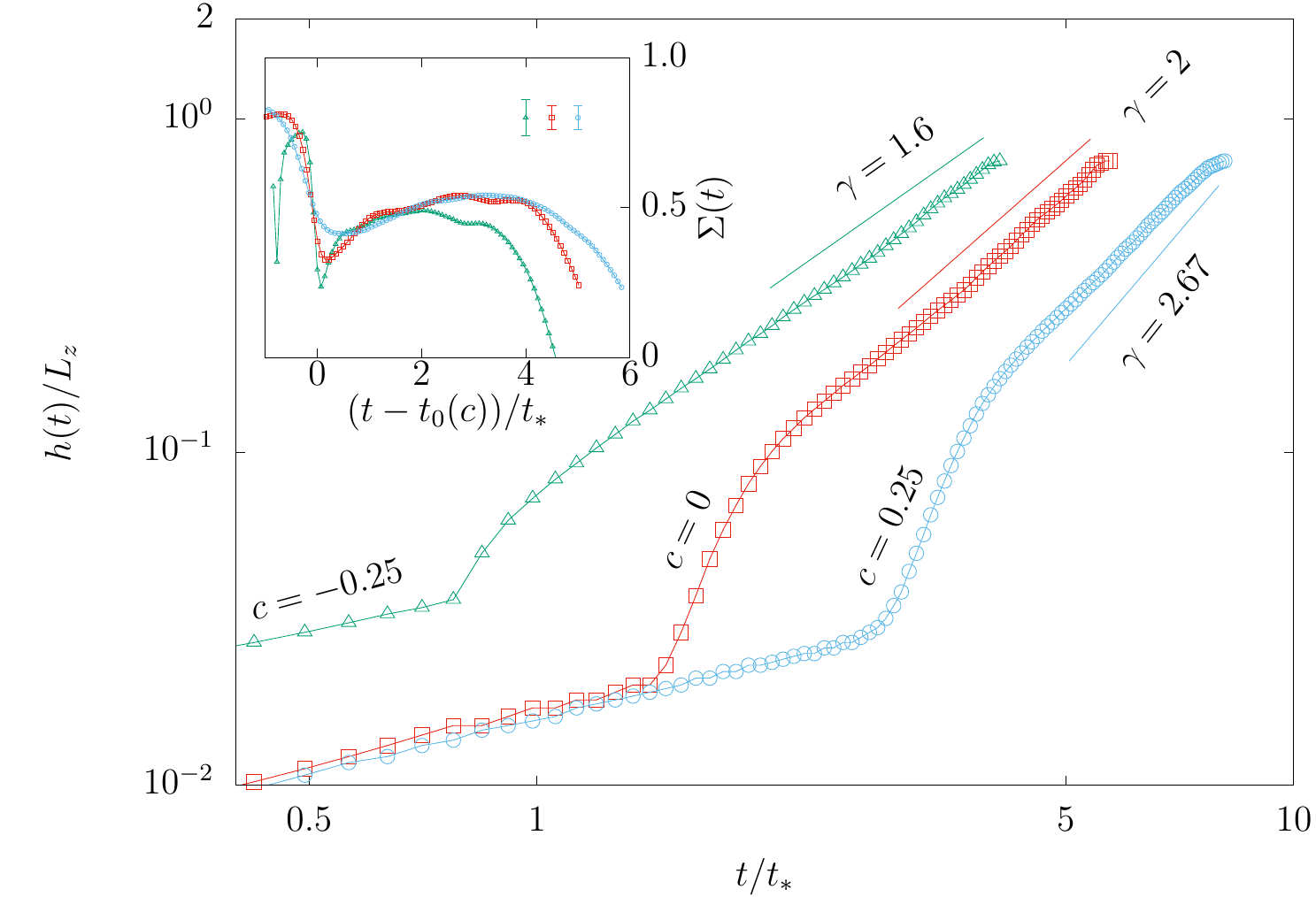}
\caption{Temporal evolution of the mixing layer $h(t)$. From left to right: $c=-0.25$ (green triangles),
$c=0$ (red squares) and $c=0.25$ (blue circles). 
The three lines represent the  power law predicted by the formula (\ref{eq2.6})
with $\gamma=2/(1-c)$. 
Inset: Efficiency of kinetic energy production $\Sigma=-(dE/dt)/(dP/dt)$ as a function of time for the three cases $c=-0.25$ (green triangles), $c=0$ (red squares) and $c=0.25$ (blue circles). The time axis is shifted by a time $t_0$ which depends on $c$ defining the onset of the self-similar growth. The bars indicate the typical amplitude of fluctuations around the mean value in the plateau region.}
\label{fig:profile}
\end{figure}

The nonlinear development of the RT instability produces a mixing zone of 
width $h(t)$. Its evolution can be determined on dimensional
grounds~\cite{cook2001transition,chertkov2003phenomenology,ristorcelli2004rayleigh} from (\ref{eq2.3}) and (\ref{eq2.1}) in the form
\begin{equation}
u(t)^2 / h(t) \simeq \beta g\theta_0(h(t)/L)^c,
\label{eq2.4}
\end{equation}
where $u(t)$ is a large-scale velocity.
Assuming that $u \simeq dh/dt$, one ends with
\begin{equation}
h(t) \simeq 
L\left(\frac{t}{t_*}\right)^{\frac{2}{1-c}},\quad
u(t) \simeq U\left(\frac{t}{t_*}\right)^{1 + c \over 1-c},
\label{eq2.6}
\end{equation}
where $U = L/t_*$ and $t_*$ was defined above.
Notice that the first expression can be reinterpreted as a standard RT diffusion 
\begin{equation}
h(t) = \alpha_c \mathcal{A}_c(t) g t^2
\label{eq:alpha} 
\end{equation}
where $\mathcal{A}_c(t) = \left(\beta \theta_0\right)^{1/(1-c)} \left(gt^2/L\right)^{c/(1-c)}$ is the time dependent Atwood number and the pre-factor $\alpha_c$ represents the generalization of the standard RT $\alpha$ coefficient \cite{alphacollaboration}. 

In order to test the above predictions, we performed direct numerical simulations (DNS) of the system of equations
(\ref{eq2.1}--\ref{eq2.2}) in a periodic domain of size $L_x \times L_y \times L_z$
 with $L_y=L_x$ and $L_z=4 L_x$ by means of a fully 
parallel pseudo-spectral code at resolution $512 \times 512 \times 2048$ 
for initial conditions (\ref{eq2.3}) with different $c$ and $L=L_z$.
For all runs we have $\beta g=1/2$, $\theta_0=1$ and $\textrm{Pr}=\nu/\kappa=1$.
RT instability is seeded by adding to the initial density 
field a white noise of amplitude $10^{-3} \theta_0$ and 
statistical quantities are averaged over $10$ independent runs.
Figure~\ref{fig:snapshot} shows examples of the vertical section of the temperature field for three different initial conditions taken at three
different times corresponding to the same width of the mixing layer 
$h(t)$.
We compute $h(t)$ on the basis of the
mean temperature profile $\overline{T}(z,t)=\int T(x,y,z,t) dx dy$ as the 
region on which $|\overline{T}(z,t)-\overline{T}(z,0)|> \delta \theta_0$ with $\delta = 5 \times 10^{-3}$
 \cite{dalziel1999self}.
In Fig.~\ref{fig:profile} we show that the evolution of $h(t)$ is in good agreement with
the power law predicted by scaling (\ref{eq:alpha})
for the three different values of $c$. A small deviation is observed for the largest
$c$ (which corresponds to the faster growth) probably because of the short
range of temporal scaling. This results confirms that the balance 
(\ref{eq2.4}) gives the correct evolution of the mixing layer, even
over non-uniform backgrounds.

From (\ref{eq2.1}-\ref{eq2.2}) we derive the energy balance equation
\begin{equation}
-{dP \over dt}=\beta g \langle w T \rangle = {dE \over dt}+\varepsilon_{\nu}
\label{eq2.6b}
\end{equation}
which defines the conversion of available potential energy 
$P(t)=-\beta g \int z \overline{T}(z,t) dz$ into turbulent kinetic energy 
$E(t) =(1/2) \langle {\bf u}({\bf r},t)^2 \rangle$. $\varepsilon_{\nu}=\nu \langle ({\bf \nabla u})^2 \rangle$
is the viscous energy dissipation and $ \langle \bullet \rangle$ represents the integral over the whole volume.
Equation (\ref{eq2.6b}) shows that not all the available potential energy is converted 
into turbulent kinetic energy. It is therefore interesting to measure the efficiency of the production of turbulent fluctuations, defined as \cite{boffetta2010statistics,boffetta2016rotating}
\begin{equation}
\Sigma = - {dE/dt \over dP/dt}
\label{eq:sigma}
\end{equation}
and to check how this is affected by the initial distribution.
The inset of Fig.~\ref{fig:profile} shows the time evolution of $\Sigma$,
which starts from a value close to $1$ unit. When the turbulent cascade develops we observe a peak in the energy dissipation which is reflected in the minimum of $\Sigma$. This occurs at a time $t_0(c)$ which depends on the initial condition and which is used to shift the different cases. In the turbulent, self-similar regime, at $t>t_0$, the efficiency of conversion of potential energy into kinetic energy reaches an almost constant plateau $\Sigma \simeq 0.5$ which is independent, within the errors, on the initial density profile.

\begin{figure}[t]
\includegraphics[width=1\columnwidth]{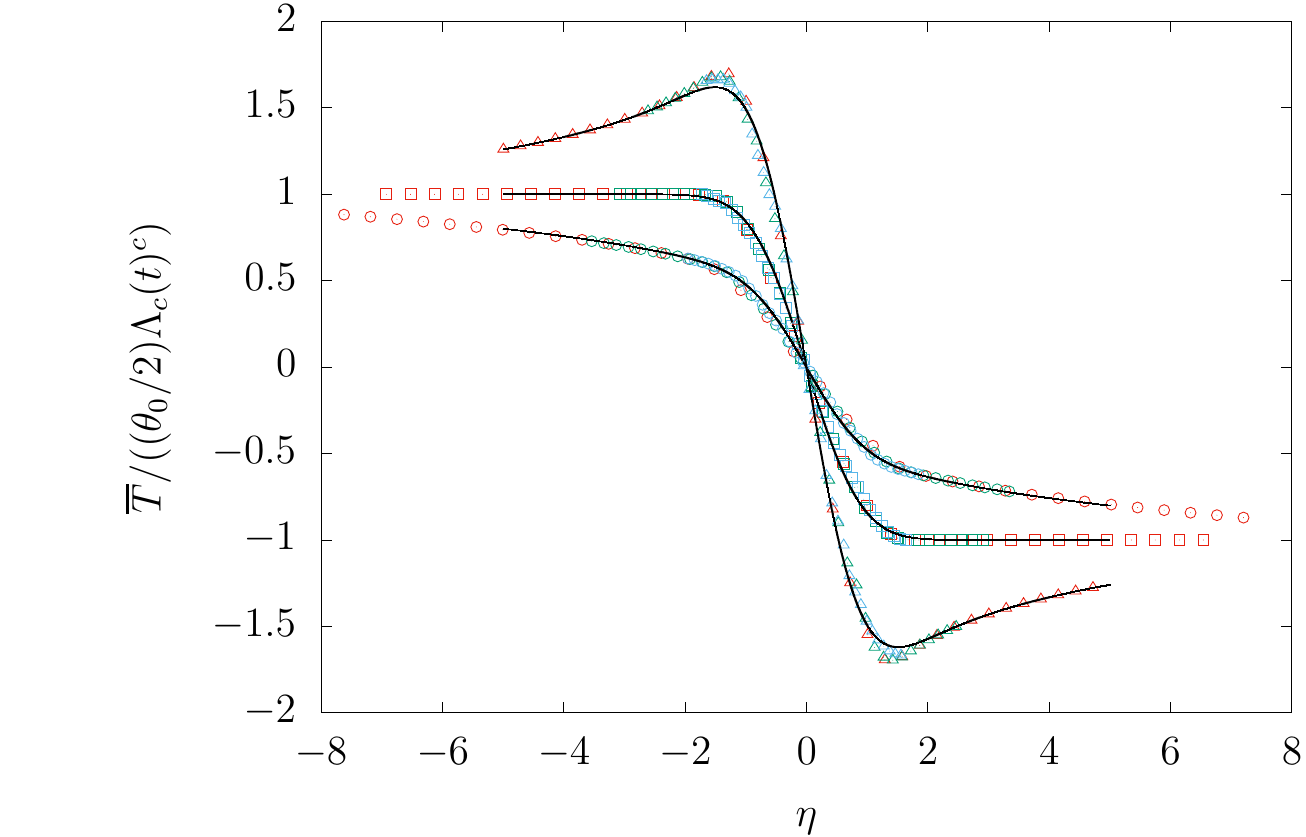}
\caption{Rescaled temperature profiles $ \overline{T}/(\theta_0 \Lambda_c(t)^c)$ averaged over ten independent runs
{\it vs} the vertical coordinate $\eta$ (\ref{eq:meanprofc}),
for $c = 0$ (squares), $c=0.25$ (circles) and $c=-0.25$ (triangles) at three different times. $\Lambda_c(t) = \sqrt{(1-c)b_c}(t/t_*)^{2/(1-c)}$ is the time scaling factor of $\eta$ in (\ref{eq:meanprofc}). The fitting parameters are: $b_0 = 6 \times 10^{-5}$, $b_{0.25} = 1.2 \times 10^{-6}$ and 
$b_{-0.25} = 7 \times 10^{-4}$. The solid lines represent the function $-|\eta|^c f_c(\eta)$, with $f_c(\eta)$ given by equation (\ref{eq:meanprofc2}).}
\label{tprofcM025}
\end{figure}

At the level of local quantities, the evolution equation for the mean temperature profile reads
\begin{equation}\label{eq:meanprof}
\partial_t \overline{T} + \partial_z \overline{w T} = \kappa \partial_{zz}^2 \overline{T}. 
\end{equation} 
Using a Prandtl Mixing Layer first-order closure 
with homogeneous eddy diffusivity $K(t)$,
the heat transfer is related to the 
local temperature gradient by  \cite{boffetta2010nonlinear}:
\begin{equation}\label{eq:closure}
\overline{w T}  
= -K(t) \left(\partial_z \overline{T}-c\overline{T}/z\right).
\end{equation} 
In the above  expression, the correction term $c\overline{T}/z$ ensures that $\overline{w T}$ vanishes outside the mixing zone, where $\overline{T}$ is given by Eq.~(\ref{eq2.3}). Neglecting the diffusive term, equation (\ref{eq:meanprof}) can be recast into 
\begin{equation} \label{eq:meanprof2}
\partial_t \overline{T} = K(t)\,\partial_{z}\left( 
\partial_z\overline{T}-c\overline{T}/z\right).
\end{equation}  
The effective diffusivity is expected to depend on time as $uh$, leading to $K(t) = b_cLU(t/t_*)^{(3+c)/(1-c)}$ with a free dimensionless parameter $b_c$. In this case a self-similar solution of (\ref{eq:meanprof2}) is obtained in the form (see the Supplementary material)
\begin{equation}\label{eq:meanprofc}
\overline{T}(z,t) = -\theta_0 \left(\frac{|z|}{L}\right)^c
f_c(\eta),\quad
\eta = \frac{z}{L}\frac{(t/t_*)^{-\frac{2}{1-c}}}{\sqrt{(1-c)b_c}},
\end{equation} 
where the function
\begin{equation}\label{eq:meanprofc2}
f_c(\eta) = \frac{2\,\mathrm{sgn}(\eta)}{\Gamma\left(\frac{1-c}{2}\right)}\int_0^{|\eta|} x^{-c} e^{-x^2}dx
\end{equation} 
is such that $f_c \to \pm 1$ as $\eta \to \pm\infty$. 
For $c = 0$ (standard RT), the solution reduces to the error function $f_0(\eta) = \mbox{erf}(\eta)$ which is known to be a good fit for standard RT evolution \cite{boffetta2010nonlinear}. 
In Fig. \ref{tprofcM025} we show that the homogeneous Prandtl approach works well also for $c \neq 0$ by plotting the rescaled temperature profiles, for the three different $c$'s considered, at three times as function of the rescaled coordinate $\eta$, as given by (\ref{eq:meanprofc}),  superposed with the solution (\ref{eq:meanprofc2}).  




\noindent {\sc Results for Shell Models.}
Because of limitation in the resolution, DNS can access the turbulent dynamics of the ML only in a limited range of scales. To get a more quantitative control of the multi-scale dynamical properties, we use a shell model for the RT instability that was  introduced in \cite{mailybaev2017}. This system defines the dynamics at discrete vertical scales (``shells'') $z_n = 2^{-n}L$ with $n = 1,2,\ldots$, where the associated variables $\omega_n$, $R_n$ and $T_n$ describe vorticity, horizontal and vertical temperature fluctuations, respectively. We modified the equations described in~\cite{mailybaev2017} by using the complex nonlinearity of the Sabra model~\cite{lvov1998}. The resulting shell model retains scaling properties of the original Boussinesq equations (\ref{eq2.1}--\ref{eq2.2}), along with some important inviscid invariants such as energy, helicity and entropy (see the Supplementary material). Having properties qualitatively similar to the full system, the shell model allows for numerical simulations in a very large range of scales, thus, serving as a natural playground for testing theoretical ideas in turbulence~\cite{biferaleARFM}.

At $t = 0$, the analogue of initial conditions (\ref{eq2.3})  must be chosen with vanishing vorticity and horizontal temperature variations $\omega_n(0)= R_n(0) = 0$, while for the vertical temperature variables we choose \begin{equation}
T_n (0) = i\theta_0 \left(\frac{z_n}{L}\right)^c
\label{eq:SM4}
\end{equation}
for all $n$. This initial condition leads to the same explosive dispersion relation $\lambda_n = \xi^{1/2} k_n^{(1-c)/2}$ as the full model (\ref{eq2.3}--\ref{eq2.2}) (see the Supplementary material). Phenomenological theory of the RT instability for the shell model is essentially identical to the one of the full 3D system~\cite{chertkov2003phenomenology}, with turbulent fluctuations propagating from small to large scales. It is convenient to characterize the size of the ML with the expression $h(t) = \sum |T_n(t)/T_n(0)-1|z_n$, which estimates the largest scale $z_n$ at which the temperature profile $T_n(t)$ deviates from its initial value $T_n(0)$. This definition is in spirit of the commonly used integral formulas for the ML width \cite{cabot2006reynolds}.

\begin{figure}
\centering
\includegraphics[width=1\columnwidth]{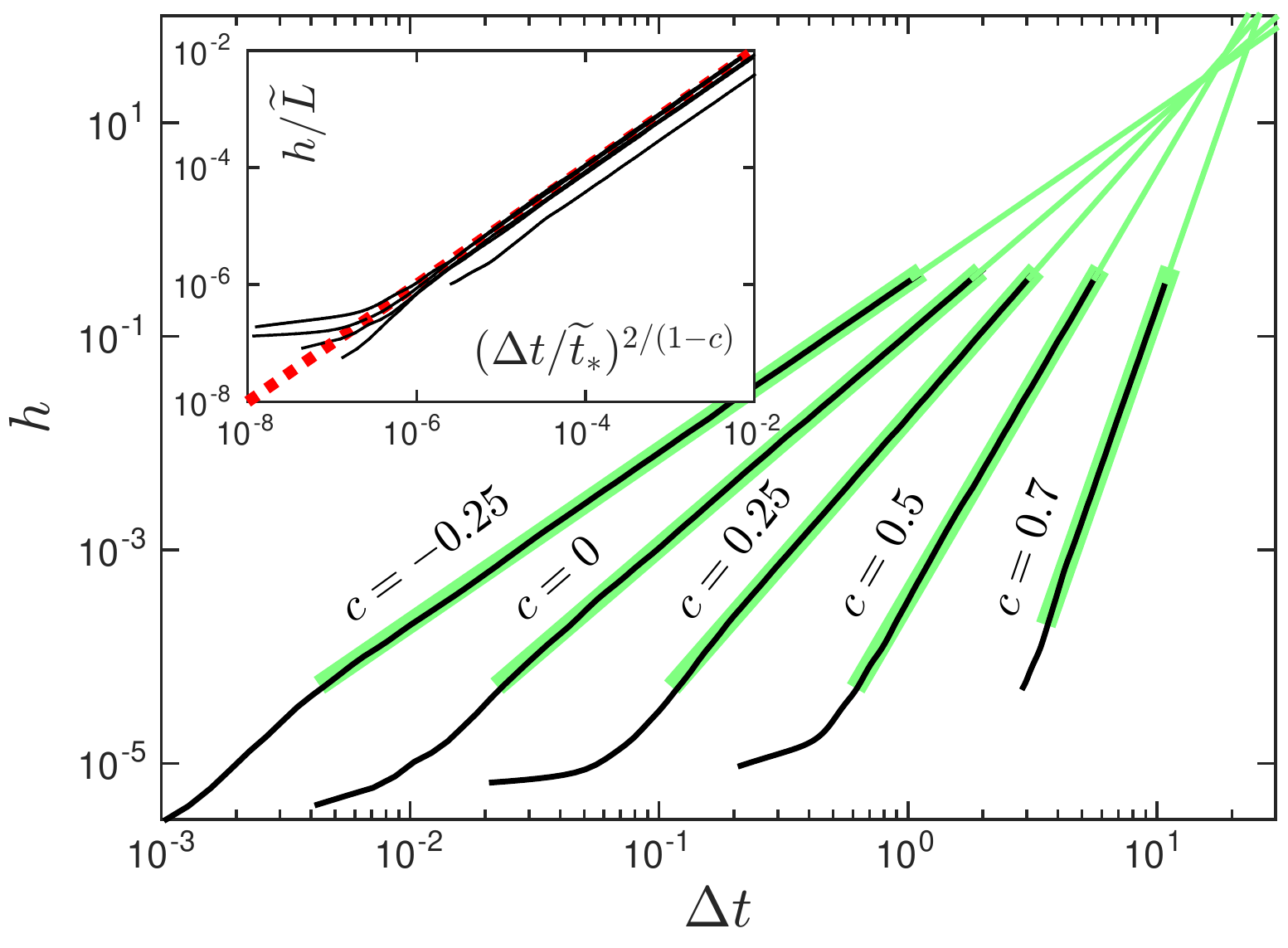}
\caption{Evolution of the ML $h$ in log-log scale for the shell model with different $c$. The parameters $L$, $\beta g$ and $\theta_0$ were set to unity. The statistics was obtained from $10^3$ evolutions, where a small random perturbation was added to the variables $R_n$ at  shells $n \ge 16$. We use $\Delta t = t-t_0$ accounting for a small initialization time $t_0$. Inset: same curves presented as  $h/\widetilde{L}$ vs. $(\Delta t/\widetilde{t}_*)^{2/(1-c)}$ and compared to the universal approximation (\ref{eq:SMUni}) shown with the dotted red line, where the larger deviation corresponds to $c = 0.7$.}
\label{figS1}
\end{figure}

By performing a large number of simulations with small dissipative coefficients and small random initial perturbations 
at small scales, we accurately verify the scaling law (\ref{eq2.6}) for $c = -0.25,\,0,\,0.25,\,0.5,\,0.7$ in Fig.~\ref{figS1}, where solid lines represent the numerical results (averaged over realizations) and the green lines  show the theoretical prediction. Here we use the small time shift $t_0$ defining the typical time for the onset of self-similar growth. 

\begin{figure}
\centering
\includegraphics[width=0.9\columnwidth]{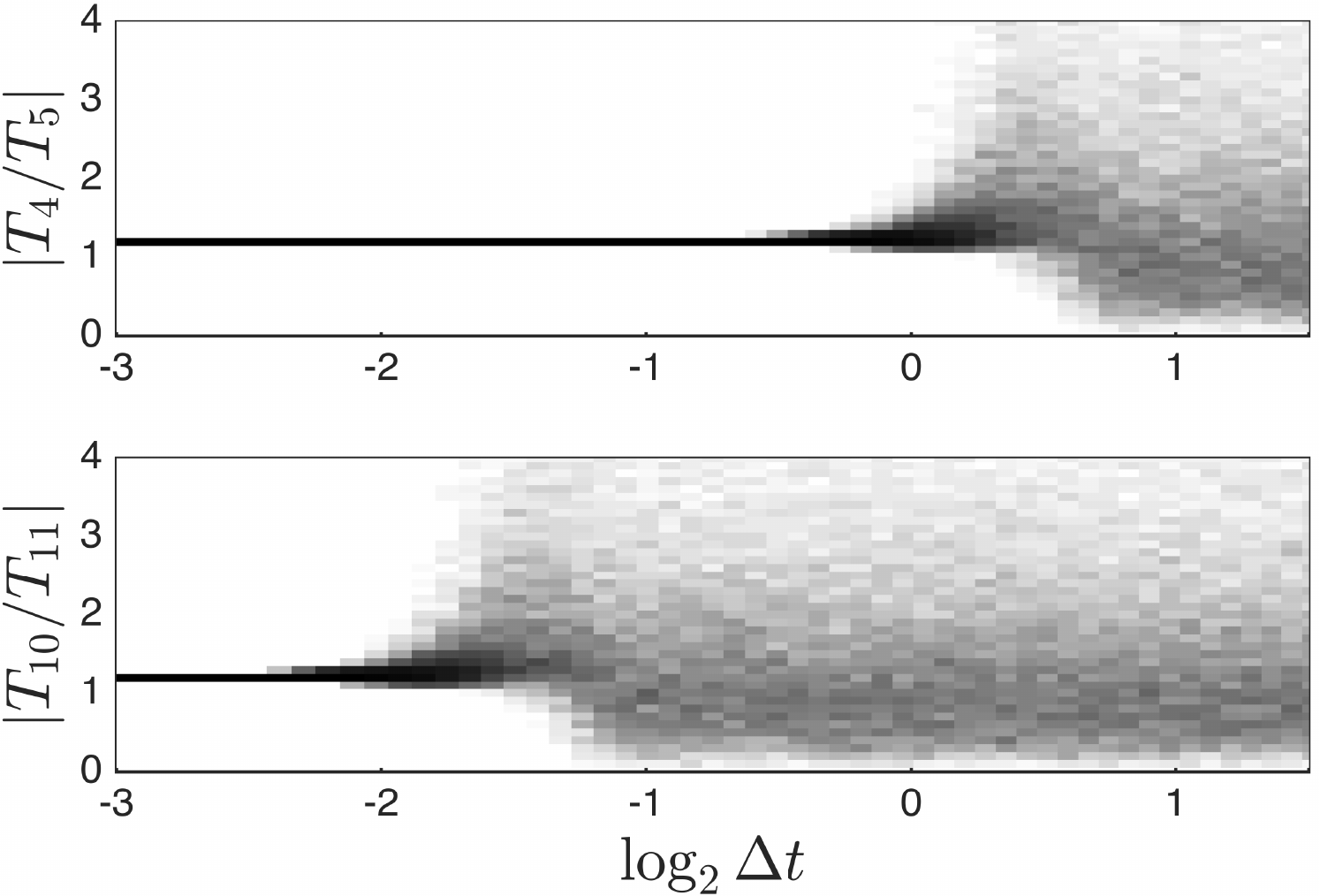}
\caption{PDFs (darker color for larger probability) for the ratios of temperature variables $|T_n/T_{n+1}|$ as functions of time: $n = 5$ (upper) and $n = 10$ (lower) panels in the case $c = 0.25$.}
\label{figS2}
\end{figure}

The results in Fig.~\ref{figS1} provide with high accuracy the dimensionless pre-factor $\alpha_c$ for the power-law growth of the ML, see Eqs.~(\ref{eq2.6}-\ref{eq:alpha}). Numerical results show that the dependence of $\alpha_c$  on the singularity exponent $c$ can be fitted well with the formula $\alpha_c \approx \alpha_L\alpha_t^{-2/(1-c)}$. The quantities $\alpha_L \approx 50$ and $\alpha_t \approx 20$ have a simple physical meaning: they redefine the  dimensional length and time scales, $\widetilde{L} = \alpha_L L$ and $\widetilde{t} = \alpha_t t_*$, which reduce the ML width expression to the universal form 
\begin{equation}
h(t) = \widetilde{L}\left(\frac{t}{\widetilde{t}_*}\right)^{2/(1-c)}. \label{eq:SMUni}
\end{equation}
This relation is validated in Fig.~\ref{figS1} (inset).
The ML reaches the size $\widetilde{L}$ at the time $\widetilde{t}$ independently of the singularity exponent $c$; this can be seen in Fig.~\ref{figS1} as an (approximately) common intersection point of the green lines. In the limit $c \to 1$ (constant temperature gradient with no singularity), the  graph $h(t)$ approaches the vertical line at time $\widetilde{t}_*$, which means that unstable modes at all scales get excited simultaneously.

It is argued~\cite{leith,eyink} that spontaneous stochastic turbulent fluctuations develop in the inverse cascade from small to large scales. In the limit of large Reynolds numbers, such behavior develops for rough (i.e., non-smooth) velocity fields, in close analogy to the $1/3$ H\"older continuity condition in the Onsager dissipation anomaly~\cite{onsager}. Non-smooth temperature profile in the RT initial conditions provides a natural rough background that can trigger similar effects in the RT turbulence. Existence of the inverse cascade of fluctuations must reflect in the stochastic growth of the mixing layer independently of the initial perturbation. 
Here, the stochastic component develops in the Eulerian evolution of velocity and temperature fields, unlike for the of turbulent Richardson dispersion where spontaneous stochasticity is predicted --and observed-- for the separation of  two Lagrangian tracers by a singular advecting velocity field~\cite{falkovich}. 

It is hard to analyze such phenomenon with the DNS due to numerical limitations. 
However, it can be conveniently studied in our shell model using the renormalized (logarithmic) space-time coordinates: 
$-n = \log_2 z_n$ and $\tau = \log_2 \Delta t$. To highlight the stochastic aspect,  we choose to measure the probability distribution function of the ratios among  temperature fluctuations at adjacent shells, $|T_n/T_{n+1}|$, which are the equivalent of velocity  multipliers used in cascade description  of fully developed turbulence~\cite{benzi,chen}.
Figure~\ref{figS2} presents the time-dependent PDFs obtained numerically in the case $c = 0.25$ starting from many initial conditions different by a very small perturbation. These results support the idea that the ML growth can be mapped to a stochastic wave in appropriate renormalized variables $(-n,\tau)$. The wave speed is constant and given by the exponent $2/(1-c)$ of ML width from Eq.~(\ref{eq2.6}). Such a wave represents a front of the turbulent fluctuations, which propagates into a deterministic left state (delta function PDF) corresponding to the initial power-law background (\ref{eq:SM4}), and leaves behind the stationary turbulent state on the right. This description naturally explains the universality of the ML evolution and its spontaneously stochastic behavior in the inertial range~\cite{mailybaev2017,mailybaev2016}. 

\noindent {\sc Conclusions.}  
We have studied numerically and analytically  Rayleigh-Taylor turbulence with general power-law 
singular initial conditions, providing  insight into situations when the mixing proceeds over a non-uniform background, e.g. in thermally stratified atmosphere.
We have shown that independently of the singularity exponent, the asymptotic self-similar growth of the ML is universal, if properly renormalized, i.e. by  looking at the mixing efficiency and at the mean rescaled Temperature profile.  We show that a closure model based on the Prandtl mixing layer approach is able to reproduce analytically the time evolution of the mean temperature profiles. By using a shell model we have provided numerical data supporting the above findings also at much larger resolution both in time and scales. This model helped to understand the behavior of prefactor in the ML growth process. Finally, we have shown that RT evolution can be reinterpreted  in terms of the phenomenon known as {\it spontaneous stochasticity}  where the growth of the mixing layer is mapped into the propagation of a wave of turbulent fluctuations on a rough background.

\noindent {\sc Acknowledgements.} L.B. acknowledges financial support  from the European Union’s Seventh Framework Programme (FP7/2007-2013) under Grant Agreement No. 339032.
A.A.M. was supported by the CNPq Grant No. 302351/2015-9. G.B. acknowledges financial support by the project CSTO162330 {\it Extreme Events in Turbulent Convection}. 
Numerical calculations have been made possible through a CINECA-INFN agreement, providing access to resources on MARCONI at CINECA.

\newpage

\section{Supplemental Material}
\subsection{Derivation of the mean profile solution, $\overline{T}(z,t)$ }
Let us write (\ref{eq:meanprof2}) as
\begin{equation} \label{eq:supp1a}
\partial_t \overline{T}
= K(t)\,\partial_{z}\left[|z|^{c} 
\partial_z\left(\frac{\overline{T}}{|z|^{c}}\right)\right].
\end{equation}  
Substituting $\overline{T}$ from (\ref{eq:meanprofc}) and dropping the common factor $-\theta_0(|z|/L)^c$ yields
\begin{equation} \label{eq:supp1}
\partial_t f_c = \frac{K(t)}{|z|^c}\,\partial_{z}\left(|z|^{c} 
\partial_zf_c\right),
\end{equation}  
where $f_c = f_c(\eta)$, with $\eta$ given by the second expression in (\ref{eq:meanprofc}) as
\begin{equation} \label{eq:supp1b}
\eta(z,t) = \frac{1}{\sqrt{(1-c)b_c}}
\frac{z}{L}
\left(\frac{t}{t_*}\right)^{-\frac{2}{1-c}}.
\end{equation}  
We can write Eq.~(\ref{eq:supp1}) in the form
\begin{equation} \label{eq:supp2}
\frac{df_c}{d\eta}\partial_t\eta
= K(t)
\left(\frac{c}{z}\frac{df_c}{d\eta}
+\frac{d^2f_c}{d\eta^2}
\partial_z\eta\right)\partial_z\eta.
\end{equation}  
Using (\ref{eq:supp1b}) and the definition of  
\begin{equation} 
\label{eq:supp2b}
K(t) = b_c \frac{L^2}{t_*}\left(\frac{t}{t_*}\right)^{(3+c)/(1-c)}
\end{equation}  
in Eq.~(\ref{eq:supp2}) leads,  after a long but elementary derivation, to:
\begin{equation} \label{eq:supp3}
\frac{d^2f_c}{d\eta^2}
+\left(\frac{c}{\eta}+2\eta\right)
\frac{df_c}{d\eta} = 0.
\end{equation}  
Denoting $g_c = df_c/d\eta$ we can recast the above expression in to:
\begin{equation} \label{eq:supp5a}
\frac{dg_c}{d\eta}
+\left(\frac{c}{\eta}
+2\eta\right)g_c
= 0.
\end{equation}  
The general solution of Eq.~(\ref{eq:supp5a}) has the form
\begin{equation} \label{eq:supp5}
g_c(\eta) = C|\eta|^{-c}e^{-\eta^2}
\end{equation}  
with an arbitrary pre-factor $C$. Finally, the solution for $f_c(\eta) = \int g_c(\eta)d\eta$ takes the form (\ref{eq:meanprofc2}), 
where $C = 2/\Gamma\left(\frac{1-c}{2}\right)$ is determined from the condition $f_c \to \pm 1$ as $\eta \to \pm\infty$.

\subsection{Shell Model for RT evolution}
We introduce the RT shell model equations in the form
\begin{equation} \label{eq:SM1}
\begin{array}{rcl}
\dot\omega_n & = & \displaystyle
-\omega_{n+2}\omega_{n+1}^*/4
    +\omega_{n+1}\omega_{n-1}^*/2
    \\[5pt]
    &&+2\omega_{n-1}\omega_{n-2}+i \beta g R_n/z_n-\nu \omega_n/z_n^2,
\end{array}
\end{equation}

\begin{equation} \label{eq:SM2}
\dot R_n =
\omega_n^*R_{n+1}-\omega_{n-1}R_{n-1}
    +\omega_n T_n^*
    -\kappa R_n/z_n^2,
\end{equation} 
\begin{equation} \label{eq:SM3}
\dot T_n =
\omega_n^*T_{n+1}-\omega_{n-1}T_{n-1}
    -\omega_n R_n^*
    -\kappa T_n/z_n^2.
\end{equation} 
This system defines the dynamics at discrete vertical scales (``shells'') $z_n = 2^{-n}L$ with $n = 1,2,\ldots$, where the associated variables $\omega_n$, $R_n$ and $T_n$ describe vorticity, horizontal and vertical temperature fluctuations, respectively. 

Equations (\ref{eq:SM1}--\ref{eq:SM3}) are analogous to those proposed in~\cite{mailybaev2017}, except for the fact that here we used the more popular Sabra model nonlinearity~\cite{lvov1998,biferaleARFM} for the vorticity Eq.~(\ref{eq:SM1}), where $\omega_n = u_n/z_n$ and $u_n$ are the velocity shell variables for the Sabra model. Notice that usually in shell model literature the equations are written using $k_n = 1/z_n$ to denotes scales in Fourier space.
Equation (\ref{eq:SM1}) without the buoyancy term has energy $E = \sum |u_n|^2$,  and the helicity $H = \sum (-1)^n|u_n \omega_n|$ as inviscid invariants in agreement with 3D Navier-Stokes equations. Equations (\ref{eq:SM2}) and (\ref{eq:SM3}) possess the inviscid invariant $S = \sum |R_n|^2+|T_n|^2$, which can be interpreted as the entropy. 

One can show that the initial condition (\ref{eq:SM4}) with vanishing $\omega_n(0) = R_n(0) = 0$ lead to the exponentially growing modes~\cite{mailybaev2017}. Let us consider small perturbations, $\Delta\omega_n$ and $\Delta R_n$, and neglect the dissipative terms. Then, Eqs.~(\ref{eq:SM1}) and (\ref{eq:SM2}) linearized near the initial state read
\begin{equation} \label{eq:SM1L}
\Delta\dot\omega_n = \frac{i \beta g}{z_n} \Delta R_n,
\quad
\Delta\dot R_n = -i\theta_0\left(\frac{z_n}{L}\right)^c\Delta\omega_n.
\end{equation}
Solution of these equations provide one unstable mode for each ``wavenumber'' $k_n = 1/z_n$ with the corresponding positive Lyapunov exponent \begin{equation} 
\label{eq:SM1LE}
\lambda_n = \xi^{1/2} k_n^{(1-c)/2},\quad
\xi = \beta g\theta_0/L^c,
\end{equation}
in the direct analogy with the RT instability of the full 3D system.

In summary, the shell model (\ref{eq:SM1}--\ref{eq:SM3}) mimics spatial variations of the vorticity and temperature fields at a wide range of scales $z_n$ in a way that closely reproduces important properties of the full RT instability. Such description can be adapted for both two and three spatial dimensions, by tuning the model coefficients to conserve the respective  invariants~\cite{mailybaev2017}. It should be stressed that the resulting models feature most phenomenological properties of the RT turbulence described by Chertkov~\cite{chertkov2003phenomenology}.

For numerical analysis, we consider dimensionless formulation with the parameters $L$, $\beta g$ and $\theta_0$ set to unity and very small dissipative parameters  $\nu = \kappa = 10^{-10}$.
We simulated numerically the model with $30$ shells. As $c \to 1$, the growth of small-scale linear modes is depleted, affecting the length of the power-law interval; see Fig.~\ref{figS1}. For example, one has $\lambda_n \propto k_n^{0.05}$ for $c = 0.9$. In this case the power-law interval is not observed unless one considers the model with a larger number of shells and much smaller dissipative parameters.

\end{document}